Ilya Solntsev[a]*, Anatoly Vorobyev[a], Elnura Irmatova[bc], Nikita Osokin[a]

**Rating evaluation of sports development efficiency using statistical analysis: evidence from Russian football**

**Abstract:**

Increasing investments into various dimensions of sports draw a significant amount of attention to the way these resources are being managed and which organizations achieve development goals with higher efficiency. This paper reviews the methodology of designing an efficiency rating model for assessing sports entities, focusing on the experience of Russian football. The authors provide an overview of various normative, statistical and econometric approaches that have been previously applied to construct rating models. The Russian Regional Efficiency of Football Development model aims to evaluate the regional federations of the Football Union of Russian via 5 dimensions: reserve training, elite sport, infrastructure, grassroots and development & promotion activities. The initial set of factors was selected by an expert panel, the finalized list of factors was derived using a multicollinearity analysis, which allowed minimizing double-counting. The scoring method of the model is based on the three-sigma rule of distribution that allowed assessing every region's performance with a 10-point scale. Support factors in the form of population density and climate were also included, since Russian regions significantly differentiate in these aspects. The findings of this paper showcased that not a single region was able to achieve a maximum 5-star rating, while regions set to host the 2018 FIFA World Cup did not score better compared to others. In conclusion the authors provide various suggestions on further developing and implementing rating models within global sports organizations.

[a] Plekhanov Russian University of Economics (Moscow, Russia)
[b] National Rating Agency of Russia
[c] Russian Presidential Academy of National Economy and Public Administration (Moscow, Russia)
* Corresponding author: ilia.solntsev@gmail.com

**Introduction**

The past decades have seen the sports industry become radically commercialized with massive investments into elite athletes, infrastructure, sports for all, disabilities sport and youth development. However, due to the variety of sports development policies adopted in various parts of the globe and the great disproportion in socio-economic, demographic, political and other factors some countries have been able to produce better results in elite sport tournaments (Vorobyev, Zarova, Solntsev, Osokin, & Zhulevich, 2016; Bernard & Busse, 2004) as well as achieve higher mass sport participation rates (Lera-López & Rapún-Gárate, 2011). This tendency is more so evident within global football, since the International Federation of Football Associations (FIFA) is one of the largest sports organizations in the world with 211 member countries. Every national football association is also faced with internal governance issues mainly linked with managing regional federations. In this case, the example of the Football Union of Russia (FUR) is most relevant, since the organization comprises 83 regional football federations, scattered across the largest country in the world.

Russia is set to host a number of prestigious football mega-events in the upcoming years: 2017 FIFA Confederations Cup, 2018 FIFA World Cup, 2020 UEFA European Championships. Russian football aims to capitalize on this once-in-a-lifetime opportunity to bring the local football development scene to the next stage. The legacy programs of the upcoming tournaments along with the nation's ambitious goals in developing the game set out clear road map for the forthcoming years. However, due to the great socio-economic, demographic and climatic disparity between all Russian regions it was impossible to apply a routine solution for developing each of the 83 regional federations within FUR. Additionally, Russia's football participation rates along with the number of qualified personnel (coaches, referees, managers) and infrastructure are well behind the levels of European counterparts (Vorobyev, Solntsev, Osokin, 2016).

This paper provides an overview of Russian football's experience in creating an efficiency rating system that was executed by the authors and approved by the FUR executive committee in May 2015.

**Related research**

In recent time, the concept of rating modelling has been significantly enhanced. Nowadays it is extremely difficult to differentiate between specific rating models due to the wide variety of methods being incorporated. On the other hand, they can be generally categorized according to the applied methodology:

1. Expert opinion models;
2. Statistical models;
3. Econometric models;
4. Mixed models.

Expert opinion modelling implies both selecting the criteria using data from interviews with experts as well as using expert opinion to account for the weights of each selected variable. The expert method can be adopted in those cases where the assessed phenomenon does not have a clear quantitative representation. For example, Hongli and Junchen (2010) applied an analytic hierarchy process (AHP) to establish the weights of their credit evaluation indices. Clinton and Lewis (2008) adopted a multirater item-response model to measure the administrative agency policy preferences within the U.S. A clear shortcoming of this approach is the reliance on the estimations of the chosen expert panel. Therefore, choosing the wrong set of experts and/or the misconception of the problem on behalf of the experts could in fact compromise the model itself and/or jeopardize the validity of the calculations.

Ratings incorporating statistical and econometric analysis are predominantly based on regression modelling. In this case, the weights for each criterion are established with the use of well-known analytical techniques. Econometric modelling is fairly common among financial market studies. Gangolf et al. (2014) combined a support vector domain description (SVDD) and a linear regression to predict credit ratings concerning the risk of arbitrary bonds. Kong-lai and Jing-Jing (2010) used a discriminant analysis and logistic regression modelling to assess the financial health of 130 listed companies in China.

Recent research in rating modelling has incorporated artificial neural networks (ANN) along with principal component analysis (PCA). PCA is usually applied for decreasing the initial dataset volume and avoiding high correlation levels between the factors (Gangolf et al., 2014; Kong-lai & Jing-Jing, 2010), additionally is could be applied to construct models base on relative PCA-attributes (Irmatova, 2016). Both methods have been used to assess sports related problems (Moura, Martins, & Cunha, 2014; Gómez et al., 2012; Schmidt, 2016). Papadimitriou and Taylor (2000) attempted to assess the effectiveness of Hellenic National Sports organizations (NSO) using a Multiple Constituency Approach. The authors used interviews with directors of each NSO to extract the constituent factors for their model. Afterwards the factors were chosen using econometric methods, such as multivariate analysis of variance and univariate analysis of variance. Similar studies were conducted for evaluating the effectiveness of sports organizations in Canada (Chelladurai & Haggerty, 1991) and Singapore (Koh-Tan, 2011).

One of the more common methods used for efficiency evaluation is the Data Envelopment Analysis (DEA). This method is based on input and output variables, which form an efficiency function via linear programming. Several studies have even tackled the problem of efficiency within the sports industry. Ahmad, Mohammad & Mohammad (2013) applied the DEA method to assess the efficiency of Iranian provincial judo committees. The authors selected 3 input variables (staff, budget and sports capitation) and 6 output variables (public sports, champions sports, sports education, sports research, sports events and active sports committees). Based on the analysis the committees were classified as efficient and inefficient. The application of this method allowed to highlight, which provinces required more resource allocation for specific development needs (i.e. more focus on sports for all activities).

A significant amount of DEA research is dedicated to measuring efficiency of professional sports clubs. Most of them are based on the Scully (1995) paradigm that implies the interrelation between sporting and financial performance. Dos Santos and de Toledo Filho (2015) measured the efficiency of football club management in Brazil. However, in this research only financial factors were involved in the DEA model itself: net assets, value of intangible assets, creditor and debtor liabilities as 'inputs' and solvency, balance sheet liquidity and profitability as 'outputs'. Subsequently, the financial efficiency (measured with DEA) was compared with on-pitch achievements of clubs (presented via league points in the corresponding season). Therefore, by combining the financial and sporting performance the authors were able to rate the clubs from most to least efficient.

Barros, Bertrand, Botti and Tainsky (2014) used a data envelopment analysis to determine the cost efficiency of French rugby clubs; Espitia-Escuer & García-Cebrián (2014, 2015) – to estimate the efficiency of Spanish football teams. However, these studies opted to incorporate both sporting and financial indicators into the DEA model. Guzmán and Morrow (2007) note that the stochastic frontier model; Cobb-Douglas production function and the input-output model are also commonly used for assessing the efficiency of sports organizations.

**Model**

According to general principles, a rating model ought to be constructed by the following guidelines:

1. Identification of factors;
2. Estimation of the initial dataset volume;
3. Selection of rating method(s);
4. Theoretical and empirical testing and subsequent moderations of the model;
5. Calculation of final results.

The factors of the model were selected by a specifically organized expert panel comprising representatives from the Football Union of Russia and regional football federations. After the criteria selection and grouping process was complete, the authors went on to test the empirical evidence to establish the model's validity and robustness.

*Factor specification*

The overall 20 criteria were clustered into 5 groups: reserve training, elite sport, infrastructure, grassroots and development & promotion activities (Table 1). The model includes both absolute value indicators as well as relative and dynamic measures. Since Russia is the largest country in the world in terms of covered territory as well as having the 9th largest population on the planet, the factor s with absolute values were to be adjusted to a region's population. The regions of the Russia Federation were clustered into 3 groups:

1. more than 2 mil. residents;
2. from 1 to 2 mil. residents;
3. less than 1 mil. residents.

No statistical analysis was required to conduct the clustering since this sort of classification is imposed by the Russian Ministry of Sport[d], the statistical reports of which were also the main source of information for this research.

The initial list of factors comprised more than 30 factors, which were eliminated after conducting a multicollinearity analysis. Since no regression modeling was used in this case, a simplified analysis was applied. Factors that showcased +/-0.7 paired correlation coefficients were subject to being eliminated from the model. However, expert opinions were also taken into account and therefore even some highly correlated factors were included in the model, since their dependence was merely statistical without any logical basis.

---

[d] Methodical recommendations on managing sports governing bodies of the Russian Federation on a regional level (In Russian). Retrieved November 6, 2016, from Ministry of Sport of the Russian Federation, http://minsport.gov.ru/function/wp-content/uploads/2014/11/Методические-рекомендации.docx

Table 1. Factors included in the Russian football efficiency-rating model

| Criteria group | Factor |
|---|---|
| *Reserve training* | Number of people registered in specialized football training centers (academies) |
| | Sufficiency of coaching staff |
| | Increase/decrease rate of people registered in specialized football training centers (academies) |
| | Results of youth teams in interregional and national competitions |
| *Elite sport* | Number of people registered in elite sports groups of specialized football training centers (academies) |
| | Football match attendance |
| | Number of players delegated to various FUR national teams |
| | Performance of professional football clubs |
| *Infrastructure* | Number of football stadiums |
| | Number of people playing on specialized fields |
| | Number of football fields and stadiums averaged by the region's population |
| | Increase/decrease rate of the number of football stadiums |
| *Grassroots* | Increase/decrease rate of people playing football |
| | Number of football players to population ratio |
| | Increase/decrease rate of sports for all degrees in football |
| | Number of sports for all degrees in football awarded in the ongoing year in proportion to the overall amount of footballers |
| *Development & promotion activities* | Federation rating |
| | Rating of regional football development centers |
| | Inclusion of football as a regional 'basic' sport |
| | Number of registered players to all footballers ratio |

'Reserve training' is set out to determine whether a regional federation is successful in providing the national football scene with quality players. The criteria group incorporates data on youth athletes and coaches. 'Elite sport' focuses on the professional side of Russian football. This group comprises factors that are relevant to a region's club representatives in national professional football competitions: top 3 male divisions, top 2 female divisions, top 2 male futsal divisions and the male beach soccer league. 'Infrastructure' addressees a region's efforts to supply footballers with the necessary amount of specialized training fields and stadiums. 'Grassroots' is dedicated to evaluating the regional federation's efforts in promoting sports for all activities through football. 'Development & promotion activities' is the most complex out of all criteria groups. The 'Federation rating' is assessed by whether or not the regional federation has a development strategy (program), accreditation within its regional sports governing body, provides timely and satisfactory statistical reports to the Football Union of Russia and keeps an up-to-date and informative website. The rating of regional football development centers is conducted by FUR according to its own specialized methodology. Inclusion of football as a regional 'basic' sport is a significant advantage for its development, since this implies additional funding and resource allocation from regional sports governing bodies.

The authors used official statistical data "1-FK and "5-FK" to extract information on 'Reserve training', 'Infrastructure' and 'Grassroots'. This data are collected by regional sports governing bodies using a unified reporting scheme approved by the Russian Ministry of Sport, Data on 'Elite sport' factors was gathered via trustworthy sports statistics websites, while 'Development & promotion activities' information was attained by interviews with representatives of regional football federations and regional sport governing bodies.

*Setting weights*

The authors chose to apply the equal weighting method for this particular model. As noted by Decancq and Lugo (2013) equal weights are most commonly applied in multidimensional indices. Among the main merits of the method, the authors highlight its simplicity and the ability to assign even importance to the factors, in cases where it is hard to clearly grade the factors by the means of their contribution to the end-result. Among the more renowned index systems using equal weighting Böhringer and Jochem (2007) acknowledge the Human Development Index (HDI), which grades income, life expectancy and education as equally important factors as well as the Ecological Footprint Index, Living Planet Index, Index of Sustainable Economic Welfare and others. The wide application of equal weighting can also be explained by the fact that they make these models easier to interpret for the public and policy makers. Since each subgroup of the model contains an equal amount of factors we were able to avoid the drawback of equal weighting mentioned by Santeramo (2015). The author explains that hierarchical indices with equally weighted factors, but with uneven subgroups could cause statistical bias in their results, since some dimensions will have a bigger impact on the overall score. However, the authors fully comprehend that the equal weighting method is far from being undisputable, as mentioned in the work of Chowdhury and Squire (2006).

We chose to not apply data-driven methods of setting weights, since they are sensitive to adding new observations or new factors to the dataset (Nardo et al., 2008). The values of the statistically objective weights become data-specific rather than factor specific. Subsequently, since yearly collected data will vary within each region, then the weights will have to be recalculated. This will jeopardize any opportunity of conducting valid trend analyses and cross-regional comparisons (Lorenz, Brauer, and Lorenz, 2016).

*Rating method*

Each of the selected factors within the model is subject to receiving a main contingent score and a support score. The overall score (estimated by combining the contingent and support scores) is then applied to the subsequent rating scale of football efficiency among Russian regions.

The scoring system in this case is conducted using statistical data on each of the factors within the model. This principal is based on the three-sigma rule. The mean of each factor is calculated from the compound dataset of Russian regions, and then the standard deviation is derived within the selected data interval. Contingent points are allocated in accordance with the deviation of the region's values from the mean of each factor. The authors used a 10-point scale, where 0 points is the worst result within the analyzed factor, while 10 points is the best result. Based on the empirical evidence, 10 data intervals were formed for the purpose of allocating contingent points (Table 2).

Table 2. Data intervals for point allocation of the Russian football efficiency rating model

| Interval | Number of points |
|---|---|
| $A \leq X_{avg} - 2\sigma$ | 0 |
| $X_{avg} - 2\sigma < A \leq X_{avg} - 1,5\sigma$ | 1 |
| $X_{avg} - 1,5\sigma < A \leq X_{avg} - \sigma$ | 2 |
| $X_{avg} - \sigma < A \leq X_{avg} - 0,5\sigma$ | 3 |
| $X_{avg} - 0,5\sigma < A \leq X_{avg}$ | 4 |
| $X_{avg} < A \leq X_{avg} + 0,5\sigma$ | 5 |
| $X_{avg} + 0,5\sigma < A \leq X_{avg} + \sigma$ | 6 |
| $X_{avg} + \sigma < A \leq X_{avg} + 1,3\sigma$ | 7 |
| $X_{avg} + 1,3\sigma < A \leq X_{avg} + 1,7\sigma$ | 8 |
| $X_{avg} + 1,7\sigma < A \leq X_{avg} + 2\sigma$ | 9 |
| $A > X_{avg} + 2\sigma$ | 10 |

The selected intervals are meant to ensure normal or approximately normal distribution of the contingent points (in the case that the normal distribution assumption of Russian regional football development data is not compromised). Figure 1 showcases that most regions receive from 4 to 5 points within a randomly selected factor, whereas only an insignificant few attain 0 (worst) and 10 (best) points.

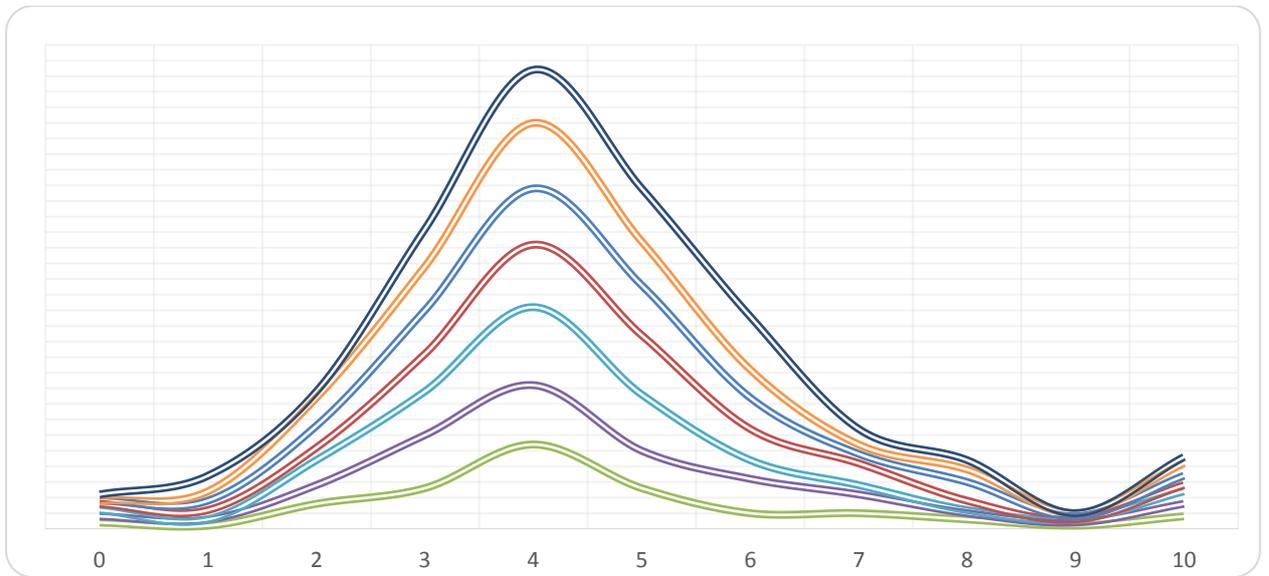

Figure 1. Example of distribution of points among Russian regions within a randomly selected factor of the football efficiency rating

The main contingent score of a *j* region is calculated by combining the weighted points received within each factor:

$$R_j = \sum_{i=1}^{I} n_{ij} w_i, \text{ where } i \in [I], j \in [N], (1)$$

*I* – the number of factors included in the model (20 factors for this model),

*N* – the number of variables (82 regions),

$n_{ij}$ – the number of points received within an *i* factor by a *j* region,

$w_i$ – the designated weight of an *i* factor,

The support scores, which are combined with the main contingent score to form the overall rating score of a region are designed to include various exogenous factors, which can seriously affect football development activities. Related literature (Hoffmann, Ging & Ramasamy, 2002) suggests that population density and climate should be included as the main support factors for this particular model. We used the average temperature of January as a proxy for the climate variable, since this time of year is usually used to evaluate the climatic disparity among Russian regions. An example of support score allocation is presented in Table 3.

Table 3. Example of support score allocation of the Russian football efficiency rating model

| GROUP | AVERAGE TEMPERATURE IN JANUARY (T), °C | POPULATION DENSITY (D), % | CORRESPONDING NUMBER OF POINTS |
|---|---|---|---|
| 2 | -15 < T ≤ -10 | 75 ≤ D | **+ 0,2** <br>**(2% of the maximum overall score)** |
| 3 | -20 < T ≤ -15 | 50 ≤ D < 75 | **+ 0,3** <br>**(3% of the maximum overall score)** |

The support scores are estimated by cross-referencing the values of a region's support factors with the corresponding number of points presented in Table 3. For example, if a region's average temperature in January equals -11 °C and its population density is 60%, then the support score will be 0.5, since its temperature value corresponds to the 2nd group (0.2 points), while its density value corresponds to the 3rd group (0.3 points).

The overall score of Russian **Regional Efficiency of Football Development** is calculated using the following equation:

$$REFD_j = R_j + D_j + T_j \text{ where, } j \in [N], \quad (2)$$

$R_j$ – the value of a $j$ region's main contingent score,

$D_j$ – the value of a $j$ region's population density support factor,

$T_j$ - the value of a $j$ region's population average temperature in January support factor.

In accordance with the acquired overall rating scores each region is classified into 1 one of the 5 football development efficiency categories (Table 4).

Table 4. Categories for the Russian football efficiency-rating model[e]

| Rating category | Overall rating score |
|---|---|
| 5 stars | ≥8 |
| 4 stars | ≥6,5 |
| 3 stars | ≥4,5 |
| 2 stars | ≥2,5 |
| 1 star | <2,5 |

**Results**

---
[e] This rating scale is designed by the Russian National Rating Agency and applied to rating models in various socio-economic spheres

Using the above-described model the authors calculated the efficiency rating of 83 regional football federations of the Football Union of Russia based on data of 2013. The top 10 regions are presented in Table 5.

Table 5. Top 10 regions according to Russian Regional Efficiency of Football Development model

| Rank | Region | Overall rating score | Rating category |
|---|---|---|---|
| 1 | Krasnodar Krai | 7,30 | **** |
| 2 | Altay Krai | 6,75 | **** |
| 3 | Moscow Oblast | 6,70 | **** |
| 4 | Republic of Mordovia | 6,55 | **** |
| 5 | Udmurt Republic | 6,15 | *** |
| 6 | Rostov Oblast | 5,95 | *** |
| 7 | Tambov Oblast | 5,90 | *** |
| 8 | Moscow (City) | 5,90 | *** |
| 9 | Volgograd Oblast | 5,75 | *** |
| 10 | Tver Oblast | 5,20 | *** |

Among the more peculiar findings presented in Table 5 is the fact that not a single region was able to acquire a 5 star rating. Such an outcome should be even more disturbing due to the upcoming football mega-events that Russia is set to host. Among the regions that will host the 2018 FIFA World Cup matches only 5 made it to the top 10: Volgograd Oblast, Moscow (City), Rostov Oblast, Republic of Mordovia and Krasnodar Krai. The other 6 regions with host-cities were ranked outside of the top 10 with a 3 star rating: Republic of Tatarstan (13th), Sverdlovsk Oblast (16th), Saint-Petersburg (24th), Samara Oblast (26th), Nizhny Novgorod Oblast (37th) and Kaliningrad Oblast (38th). 43 regions received a 3 star rating, 30 regions received a 2 star rating and only 5 regions were given a 1 star rating.

**Conclusion**

The authors were able to construct an efficiency rating model to evaluate the level of football development of regional football federations within the Football Union of Russia. The evidence from Russian football shows that efficiency evaluation could be regarded as one of the key components of strategic decision-making for sports governing bodies, since it provides a descriptive layout of structural strengths and weaknesses. One of the main challenges of constructing such efficiency rating models is the absence of both theoretical and empirical retrospect. Such a fact forced the authors to adapt general statistical and econometric approaches to the specifics of football development.

This model also allows to cross-analyze the trends of regional football development not only on a national scale, but also on an interregional level. The Football Union of Russia currently comprises 10 interregional football associations, which are compiled mainly through a geographical principle. This means that the members of these associations will most likely possess similar demographic and climatic attributes. Therefore, the efficiency rating model will allow benchmarking the best national and interregional football development practices, which can lead to designing specific blueprints.

The authors feel that global sports will require scientifically-based tools to objectify strategic decision-making and budget allocation. For example, FIFA aims to initiate its new FORWARD development programs where each of the 211 national FAs will annually receive 5 mil USD, which will result in more than 1 bill USD of additional expenditures per year. Such an approach is admirable and risky at the same time, since it denotes the efforts of efficient FAs, which are able to capitalize on the resources they are allocated, whilst also pardoning the relative mismanagement of others. This can be further implemented by continental football federations, since they also promote development policies. The logic of signaling out better performing associations could also be transferred to the distribution of power among members: more voting rights, presence in deferent committees, etc.

This paper is aimed at providing an example of how an efficiency rating model for sport organizations could be designed. However the authors fully comprehend that the specifics of the rating methodology (selection of criteria, setting of weights, inclusion of support factors, etc.) are all subject for discussion and should be modified in accordance with the needs and strategic goals of the organizations that wish to implement such models.

**Acknowledgement**